# Opportunities for Technosignature Science in the Planetary Science and Astrobiology Decadal Survey

A report from the First Penn State SETI Symposium

Jacob Haqq-Misra (Blue Marble Space Institute of Science)
Reza Ashtari (Johns Hopkins University)
James Benford (Microwave Sciences)
Jonathan Carroll-Nellenback (University of Rochester)
Niklas A. Döbler (University of Bamberg)
Wael Farah (SETI Institute)
Thomas J. Fauchez (NASA GSFC, American University)
Vishal Gajjar (University of California: Berkeley)
David Grinspoon (Planetary Science Institute)
Advait Huggahalli (HGHS Science Research)
Ravi K. Kopparapu (NASA Goddard)
Joseph Lazio (Jet Propulsion Laboratory)
George Profitiliotis (National Technical University of Athens)
Evan L. Sneed (Penn State University)
Savin Shynu Varghese (SETI Institute)
Clément Vidal (University of California: Berkeley)

Solar system exploration provides numerous possibilities for advancing technosignature science. The search for life in the solar system includes missions designed to search for evidence of biosignatures on other planetary bodies, but many missions could also attempt to search for and constrain the presence of technology within the solar system. Technosignatures and biosignatures represent complementary approaches toward searching for evidence of life in our solar neighborhood, and beyond. This report summarizes the potential technosignature opportunities within ongoing solar system exploration and the recommendations of the "Origins, Worlds, and Life" Planetary Science and Astrobiology Decadal Survey. We discuss opportunities for constraining the prevalence of technosignatures within the solar system using current or future missions at negligible additional cost, and we present a preliminary assessment of gaps that may exist in the search for technosignatures within the solar system.

**1. Introduction**

The three main trends of the modern search for technosignatures were proposed over six decades ago. Cocconi and Morrison (1959) first suggested the search for communicative technosignatures, which led to the development of numerous SETI (search for extraterrestrial intelligence) programs worldwide to search for such signals at radio and optical wavelengths that could be emanating from planets around nearby stars. Dyson (1960) proposed a search for non-communicative technosignatures by searching for evidence of astrophysical-scale megastructures in exoplanetary systems, which could be identifiable through infrared excesses. Bracewell (1960) presented an alternative to SETI searches around other stars and instead suggested a search for extraterrestrial artifacts (SETA) that could be present in the solar system.

However, only a handful of technosignature searches have been conducted within the solar system, which have consisted of searching for objects at stable Lagrange points (Freitas and Valdes 1980; Valdes and Freitas 1983), searching for artifacts on the lunar surface (Lesnikowski et al. 2020), conducting radio observations of the interstellar object 'Oumuamua (Enriquez et al. 2018; Park et al. 2018; Tingay et al. 2018; Harp et al. 2019), and examining sky surveys for anomalous transients (Villarroel et al. 2021). With such a limited number of searches conducted so far, the prevalence of technosignatures in the solar system remains very poorly constrained (Haqq-Misra and Kopparapu 2012). Our goal in this report is to show that the search for past or present technosignatures in the solar system falls within the capabilities of current and planned planetary science missions.

One core objective of SETA is to exclude the presence of any technological structures larger than 10m on bodies within the Solar System having solid surfaces. This size is a starting point, but comparable to the size of spacecraft that we have sent into interstellar space. Such a goal is achievable through visible or near-IR images of a target body with a spatial resolution not exceeding 3 m/pixel. A second objective is to exclude anomalous waste heat from artifacts, as it is an inevitable side-effect of work being performed by an active technological artifact. Such a goal is achievable through IR measurements on target bodies. Such objectives are clearly falsifiable (given reasonable assumptions about the recognizable visible qualities of artifacts) and in the future, more and more stringent limits could be placed on the existence of smaller and smaller sized artifacts, or lower and lower amounts of waste heat as our instruments become more sensitive and our solar-system exploration more systematic.

It is worth noting that SETA is an exploration of nearby places through deep time. That is, looking for artifacts that may have been here for millions of years, so it samples the long-term history of possible visits to the solar system. This proximity offers opportunities to resolve ambiguities and uncertainties of candidate detections with further missions, such as return-sample missions. While traditional SETI is generally a passive activity which involves listening to stars or galaxies for anomalous signals, SETA involves active exploration that looks at nearby objects and eventually send probes to them. Solar system SETA thus requires a different skill set, which has much more overlap with the methods used by the planetary science community. Current and planned solar system missions could place stronger constraints on the prevalence of technosignatures, in many cases without any additional modification or cost to the mission. What is more, this SETA effort could be part of a planetary defense program, as technology from extraterrestrial intelligence constitutes another set of potential threats next to natural threats (e.g. asteroids and comets) or contamination threats from microbial life-forms that could affect planet Earth in one way or another.

This report summarizes the potential technosignature opportunities within the solar system, which includes the recommendations of the "Origins, Worlds, and Life" Planetary Science and Astrobiology Decadal Survey. We note that this planetary science decadal survey explicitly mentions that "*The scientific identification and initial validation of technosignatures is in scope but the application of such signatures in survey studies is out of scope.*" The objective of our present report is to demonstrate the relevance of technosignature science to many existing and planned solar system missions and encourage the planetary science community to consider including technosignature science as part of the stated scientific justification in mission planning. The 2020 Astronomy and Astrophysics decadal report considered *"Life's global impacts on a planet's atmosphere, surface, and temporal behavior may… manifest as potentially detectable exoplanet biosignatures, or technosignatures—if that life is technologically capable."* Technosignature science, thus, is an inherent part of the search for life elsewhere. **We also urge planetary scientists to consider the possibility that evidence of technosignatures could be detected in a wide range of data sets as an unexplainable anomaly, so it is worth being aware of the possibility of such an event, even if the likelihood may be unconstrained**.

We summarize the opportunities and gaps for solar system technosignature science in the table below (click here to view as a spreadsheet). The remaining sections briefly discuss capabilities for detecting each class of technosignature.

| Solar System Technosignatures | | Non-terrestrial Artifacts | | | Evidence of Industrial Activities | | | Effects of Interstellar Travel | | | |
|---|---|---|---|---|---|---|---|---|---|---|---|
| | | Surface Artifacts | Lurkers | Interstellar Artifacts | Mining | Waste Energy/Materials | Geochemical Anomalies | Interstellar Drive | Laser Propulsion | Gravitational Anomalies | Atmospheric Impact/Re-entry |
| Orbiters, Rovers, and Probes | Mars | | | | | | | | | | |
| | Moon | | | | | | | | | | |
| | Sun | | | | | | | | | | |
| | Mercury | | | | | | | | | | |
| | Venus | | | | | | | | | | |
| | Uranus probe/orbiter | | | | | | | | | | |
| | Ceres | | | | | | | | | | |
| | Enceledus | | | | | | | | | | |
| | Europa Clipper | | | | | | | | | | |
| | Titan Dragonfly | | | | | | | | | | |
| Small Bodies | Psyche asteroid mission | | | | | | | | | | |
| | Asteroid surveys | | | | | | | | | | |
| | Lucy | | | | | | | | | | |
| | KBO surveys | | | | | | | | | | |
| | Planetary defense (NEO Surveys) | | | | | | | | | | |
| | Sample return | | | | | | | | | | |
| | Interstellar Object Intercept | | | | | | | | | | |
| Observatories | Radio / ALMA | | | | | | | | | | |
| | Chandra | | | | | | | | | | |
| | LUVEx | | | | | | | | | | |
| | Rubin | | | | | | | | | | |
| | JWST | | | | | | | | | | |
| | Roman | | | | | | | | | | |
| | High energy particle detectors | | | | | | | | | | |
| | Extremely Large Telescopes | | | | | | | | | | |
| Earth | Human exploration | | | | | | | | | | |
| | Earth analog/Silurian hypothesis | | | | | | | | | | |

Legend: YES / MAYBE / NO

## 2. Non-terrestrial Artifacts

This class of technosignatures refers to any extraterrestrial technological artifact that resides within the solar system. This includes surface artifacts that could be present on planetary surfaces, long term "lurker" artifacts residing in nearby space, and interstellar objects passing through the solar system.

*2.1 Surface artifacts*

Artifacts on a planetary surface or subsurface could be detectable with orbiters, rovers, or probes that are exploring particular planetary bodies. Such artifacts could be functional or defunct. Functional artifacts could be identified through optical albedo measurements or infrared waste energy and may also emit detectable communication signals. Both functional and defunct artifacts could be identified using radar or by analysis of surface images.

The Moon is the nearest object we can observe. We have had the Lunar Reconnaissance Orbiter (LRO) in low orbit around the Moon since 2009. It has taken >2 million images at high sub meter resolution and a few sites at resolutions of 0.5 m. This kind of resolution enables us to observe even Neil Armstrong's boot prints. However, the vast majority of the photos have not been inspected by the human eye. Searching these millions of photographs for non-terrestrial artifacts (NTAs) would require an automatic processing system. Development of such artificial intelligence (AI) is a low-cost initial activity for finding artifacts on the Moon as well as near Earth Trojans or the co-orbitals. Examination of lunar images has been discussed by Davies (Davies and Wagner, 2013). An unsupervised machine learning approach for this search has been developed by Lesnikowski et al. (2020). Note that this approach has been vindicated by the recent AI analysis of 2 million images from LRO, which revealed rock falls over many regions of the Moon (Bickel et al., 2020). Such an approach can verify the existence of known technology on the moon, such as some of the Apollo Lunar Modules impact sites, and can reveal anomalous features that may not have obvious explanations. Similar approaches to detecting surface anomalies could be conducted on Mars with HiRISE and on other planetary surfaces when high-resolution images are available.

The Cassini mission had a radar imager to survey the surface of Titan through its thick atmosphere. The radar imager operating at 13.7 GHz made the first detailed surface map of Titan with a pixel resolution of 0.3 km (Porco et al. 2005). Radar imaging, or "radiometry", on-board Cassini, has also allowed determination of the temperature of any object. Future solar system exploration missions with radar imaging capabilities would be ideal to find artificial surface or subsurface artifacts on solar system bodies. For example, the Chang'e 3 lunar rover includes ground-penetrating radar to search for subsurface water ice, brines, and lava tubes, which also may be able to constrain the presence of subsurface artifacts on the moon.

The lifetime of surface artifacts that are no longer active will be limited by the resurfacing timescale on each planetary body. Many of the rocky planets and moons have undergone extensive resurfacing after forming their initial crustal layers, limiting the timespan in which artifacts can remain on the surface. Wind is a likely form of weathering for any surface with an atmosphere, having been observed on Mars, Venus, and Titan. Corrosion in highly acidic environments such as Venus likely also limits artifact lifetimes. Erosion occurs on surfaces with liquids, such as Earth and Titan. Both Mars and Venus may have once been tectonically active; Mars is no longer tectonically active, while the overall level of tectonic activity on Venus is highly unconstrained. However, mapping of both suggests major resurfacing events have occurred from impacts and/or mantle plume volcanism. The Jovian and Saturnian moons frequently undergo massive resurfacing events due to tidal flexing and heating. Searches on such bodies for surface artifacts can only constrain the presence of technology in recent solar system history only as far back as the resurfacing timescale.

*2.2 Lurkers*

Active or remnant probes in the solar system are sometimes referred to as "lurkers" (Bracewell 1960, Papagiannis 1978, Brin 1999) and could be plausibly found in several locations. These include the gravitationally stable Lagrange points of the various bodies in the solar system, but also privileged orbits around Earth or planets, such as the horseshoe, tadpole, or quasi-satellite orbits (Benford 2019). These "co-orbital" objects approach Earth very closely and annually at distances much shorter than anything except the Moon. They have the same orbital period as Earth. These objects are assumed to be natural, such as asteroids which have wandered into this type of orbit for a period of centuries to millennia.

Radar systems are one of the primary methods for detecting such small objects within our solar system. Often used to study planets and asteroids in our solar system, radar can also be used to look for lurkers in our solar system. The co-orbital objects have not been pinged or imaged by any planetary radar yet. Nearby co-orbitals are detectable with radar. In any case, radar can ping the objects, meaning that a signal reaches them, but the return signal may be too weak to detect at Earth. If there is an extraterrestrial probe there, it might sense that it had been noticed by us. Such activity could even provoke a reply. This possibility was suggested by Tough (1998) and Vakoch (2011), who argued that possible resident lurkers in the solar system may be waiting for Earth to initiate active communication, such as by transmitting a radio signal in the direction of such an artifact.

Another method to search for lurkers is direct imaging using planetary radars. The Arecibo telescope used to host the world's most powerful radar transmitter, capable of up to ~900 kW of output power operating at S-band (2380 MHz). It also had a pulsed P-band (430 MHz, 70 cm) radar transmitter with an effective output power of the order of ~100 kW. Although Arecibo is no longer operational, Deep Space Network antennas across the world are currently being used for planetary radar projects to look at near-Earth objects. Radars with higher power, longer baselines and shorter wavelengths will provide higher-resolution imaging capability, beneficial for detecting and identifying various lurker dimensions and surface features. Future radar observations such as these can help to constrain the prevalence of lurkers at stable orbits in the solar system.

The search for lurkers can also harness passive observations at optical, infrared, and radio wavelengths. Such observations could be performed on known near-Earth objects as well as at known stable orbits (Benford 2019). Optical observations could help to identify anomalous morphological or reflective features, while infrared observations could identify excesses due to waste energy from a functional artifact. Targeted radio observations could also be used to identify communicative signals or reflected terrestrial radio frequency interference from artifacts. Some radio telescopes also conduct all-sky imaging looking for transient phenomenon, such as the Long Wavelength Array radio telescope that operates from 10-90 MHz, which could be compared with the positions of known satellites to identify anomalies.

*2.3 Interstellar Artifacts*

Objects originating from another planetary system that pass through the solar system are another possibility to examine for evidence of technology. Interstellar artifacts could be functional exploratory spacecraft, similar to the Voyager missions, as well as technological debris or waste that is drifting through interstellar space. Distant visitors like 'Oumuamua are examples of interstellar objects that provide opportunities to learn how to best investigate transient phenomena passing through the solar system. The prevalence of objects like 'Oumuamua is only loosely constrained (Levine et al., 2021), so the continued study of similar interstellar objects can also enable the development of methods that could be used to identify, track, and study interstellar artifacts.

Solar system missions that study small bodies such as the Lucy and Psyche NASA missions have instrument characteristics (spatial, temporal resolution, sensitivity, etc.) tuned to detect and eventually characterize small artifacts.

From Earth, dozens of Kuiper Belt Object (KBO) surveys (Kavelaarset al., 2008) have been performed over the last decades using a large variety of ground based observatories building an increasingly large catalog. An improved demographic of KBOs helps to identify potential outliers and anomalies that could be interstellar artifacts.

JWST and even future space missions such as the Astro2020 decadal survey mission concept LUVOIR (now LUVEX since the LUVOIR mission concept has been combined with the HabEX mission concept) have an extensive planned solar system observation program. LUVEX specifically will have an unprecedented spatial resolution that would contribute to image outer solar system objects such as Europa or Pluto with resolution comparable to space probes dedicated to close-in observation such as New Horizons.

Several missions are planned to intercept interstellar objects passing through the solar system. For instance the project Lyra from the Initiative for Interstellar Studies (i4is) assessing the feasibility of a mission to 'Oumuamua (Hibberd et al., 2022) and the Comet Interceptor spacecraft ESA, JAXA joint mission planned to be launched in 2029. The Comet Interceptor (Jones and Snodgrass 2019) will be positioned at the Earth-Sun Lagrange L2 point waiting for long period comets to intercept. A dedicated intercept mission could also remain on standby to launch on short notice to obtain close observations of an incoming interstellar object.

Some radio observatories and interferometers, such as the GBT, ATA, VLBA, and VLA, are able to detect our own interstellar travelers with high precision. For example, the VLA and ATA can provide arcsecond resolution and the VLBA can provide milliarcsecond resolution for characterizing interstellar objects. Observations and models can improve understanding of the expected velocities of interstellar objects (Mamajek 2017).

## 3. Evidence of Industrial Activities

Activities by extraterrestrial visitors within the solar system could leave long-lasting traces that would otherwise be anomalous. This is often discussed within the framework of self-replicating probes that previously visited or are currently resident in the solar system (Ellery 2022). Possibilities for extraterrestrial activities that could be detectable include evidence of mining asteroids or planetary surfaces, waste energy or materials on planetary surfaces that are the result of industrial or other technological processes, and geochemical anomalies within rocks that reveal ancient traces of technology.

### 3.1 Mining

An extraterrestrial presence in the solar system could find the asteroid belt to be useful as a source of raw materials. Searching for forensic evidence of asteroid mining has been suggested by Papagiannis (1978, 1983), which could place limits on the presence of active or defunct extraterrestrial industrial activities. Evidence of active extraterrestrial mining on asteroids and terrestrial planets could be detectable through anomalous infrared excesses that originate from manufacturing processes. Evidence of prior extraterrestrial mining could be detected from visual morphological features that could be observed by orbiters, rovers, and probes on specific planetary bodies as well as by ground- and space-based telescopes.

### 3.2 Waste Energy/Materials

Extraterrestrial industrial activities on asteroids or terrestrial planets may leave detectable waste materials. The refuse from industrial processes could take the form of structures and other materials that were left behind as waste, much like the Apollo astronauts left significant equipment on the lunar surface, which could be observable by orbiters, rovers, and probes on specific planetary bodies as well as by ground- and space-based telescopes. Davies and Wagner (2013) suggested that nuclear waste, such as spent radioactive fuel, would be a compelling technosignature if discovered on the surface of a planetary body, which could be characterized by missions capable of placing limits on gamma ray emissions.

### 3.3 Geochemical anomalies

Technological activities that have occurred within the solar system could cause geochemical anomalies that would be discernible from other non-technological geological formations. The recognition of the Anthropocene as a unique geological epoch is motivated in part by anthropogenic changes to the rock record (e.g., Lewis and Maslin 2015), which will provide a lasting record of human technology on Earth even in the event of human extinction. One example is the prevalence of concrete as a building material, which could exert a uniquely detectable geologic signature far into the future. Microplastics are another example of a geological signature that could persist for long durations. Other industrial processes on Earth also disrupt sedimentary processes that contribute to the unique stratigraphic layer of the Anthropocene (Zalasiewicz et al. 2013, Waters et al. 2016). Evidence of extraterrestrial technology could likewise take the form of geochemical anomalies on planetary surfaces, which would exhibit chemical compositions or molecular structures that are only known to occur through technological processes or of unknown origin. Such technosignatures could be investigated by orbiters, rovers, and probes that are capable of mineralogical analysis. A sample return mission would provide an ideal opportunity to search for such anomalies. Some remote detection of geochemical anomalies could be possible, depending on the characteristics and extent of the anomalous features.

The geological study of the Anthropocene on Earth provides a way to test the idea of searching for geochemical anomalies on other planetary bodies. A further, perhaps more speculative, possibility is the Silurian hypothesis, in which a previous Earth-based technological civilization existed prior to the evolution of hominids (Schmidt and Frank 2019). Further exploration and characterization of Earth's geology and geochemistry is one way to provide constraints on the Silurian hypothesis.

**4. Effects of Interstellar Travel**

Extraterrestrial travel itself could exert a detectable signal, which would depend on the method of propulsion. Possibilities for detectable technosignatures associated with interstellar travel are interstellar drive, laser propulsion, gravitational anomalies, and atmospheric re-entry/impact.

*4.1 Interstellar drive*

Interstellar ships of significant mass would have drive plumes that might be detectable given the size and distance. The observational characteristics of the engines would depend on the class of engine. Ships that used nuclear fusion pulses, like Project Orion (Schmidt et al. 2022), might yield hard X-ray flashes (although this may be doppler-shifted depending on the speed of the ship). An alternative means of propulsion might involve nuclear fusion-powered ion engines, which would primarily emit thermal radiation that peaks in the ultraviolet or visible.

Interstellar ships moving towards us and accelerating/decelerating while traveling near the speed of light would also have the emission from their drive plumes potentially amplified via relativistic beaming. Interstellar ships may also communicate with their home system or with each other, and such communication could be intercepted.

*4.2 Laser propulsion*

Extraterrestrial spacecraft that are accelerated by laser propulsion toward the solar system could conceivably be detected by observing the laser pulses themselves. The current best model of how laser-propelled probes can be accelerated involves a high-throughput phased-array laser that is used for momentum transfer to achieve the required G-force, but only during the very first few seconds to minutes of the mission (Parkin 2018). After this initial pulse, the laser beam exceeds the sail diameter and efficiency declines; at this point, the beam missing the sail will be observable over interstellar distances. Such extremely bright beams could be detected as a transient in SETI searches of the near galaxy such as Breakthrough Listen (Benford and Benford 2016). Microwave frequencies can be used for power beams as well.

Breakthrough Starshot plans to send hundreds to thousands of probes to the Alpha Centauri system. For similar missions,depending on the duration between each launch, some probes may have already arrived at their destination while other probes continue to be sent. Such a model suggests that the detection of laser pulses in the direction of the solar system could be evidence of propulsion, although laser pulses may also be used for other purposes, such as communication.

### *4.3 Gravitational anomalies*

A large extraterrestrial spacecraft could conceivably exert detectable gravitational anomalies on Kuiper belt objects, in the Oort Cloud, or on other objects as it approaches the solar system. One way to constrain the prevalence of such technosignatures is to search for anomalies in the orbital dynamics of Kuiper belt objects and other objects with known trajectories.

A massive non-radiating object also could reveal its gravitational fingerprint by acting as a gravitational lens. One consideration is the gravitational focal distance of the object as a function of its mass. For reference, the gravitational focal distance of an object the mass of the sun is 550 AU. If an object with the right mass happens to be one focal distance away from earth, gravitational lensing of background stars and galaxies might reveal its presence. Gillon (2014) and Tusay and Huston et al. (2022) further suggested that gravitational lens communication could be used by nearby probes, which could be detected by looking for solar gravitational lens anomalies around the solar focal point.

### *4.4 Atmospheric impact/re-entry*

The entry or re-entry of an extraterrestrial spacecraft into a planet's atmosphere, or its impact on a planetary surface, could be detectable by high resolution imagers in orbit around a planetary object or on its surface. Such observations would be detected as anomalous transient phenomena within a planet's atmosphere (for re-entry) or on the surface (for an impact).

The atmospheric re-entry of spacecraft is similar to meteors entering Earth's atmosphere. In both cases significant ablation produces bright phenomena at optical wavelengths. Observations of meteors with the Long Wavelength Array telescope at low frequencies have found radio emissions produced by the plasma due to this ablation (Obenberger et al. 2014). Most of these meteor events are short-duration events (less than a second), but the radio emission persists for a couple of minutes. No radio emissions have yet been observed for spacecraft entry, although long-duration optical events have been observed for spacecraft reentry that are similar to meteor showers (lasting a couple of minutes). This technique could be applied to study atmospheric re-entry in general, although the signal will be fainter for more distant planets.

Ongoing Moon and Mars exploration programs can place some constraints on atmospheric impact or re-entry technosignatures. The Uranus orbiter/probe recommendation could place constraints on re-entry during the duration of the mission. Archival searches may also be possible using data from the Cassini/Huygens mission as well as the Juno mission.

## 5. Recommendations

The summary of technosignature science capabilities outlined in this report is intended to highlight the opportunities for constraining the prevalence of technosignatures through ongoing solar system exploration and future missions. Studies of nearby inner solar system objects such as the Moon, Lagrange point objects and Earth co-orbitals can be done independently. The easiest low-cost initial activity for finding alien artifacts is to inspect the many LRO images of the moon. Technosignature searches can be conducted commensally with other observations and do not require additional cost. Further exploration of the solar system will continue to yield new opportunities to constrain the prevalence of extraterrestrial technology.

This report recommends that solar system missions in planning or design phases consider the possible technosignature science that could be conducted with the planned mission architecture, and we encourage mission teams to include technosignature science as one of the stated science objectives. Solar system missions in data analysis phases should also consider possible constraints that could be placed on technosignatures from planned or ancillary analyses. This is not a call for technosignature science to drive solar system exploration, but instead this is a call to consider including technosignatures as ancillary science. As astrobiology seeks to understand the origin, distribution, and future of life in the universe, it is worth utilizing the vast array of solar system exploration missions to better constrain the prevalence of extraterrestrial technology in our solar neighborhood.